\documentclass[11pt]{article}

\usepackage[margin=1in]{geometry}

\usepackage{times}

\usepackage{amsmath,amssymb,amsfonts}

\usepackage{graphicx}
\usepackage{booktabs}
\usepackage{caption}
\usepackage{subcaption}

\usepackage[hidelinks]{hyperref}

\usepackage{needspace}

\begin{document}

\title{Applying Embedding-Based Retrieval to Airbnb Search}
\date{Jan 10th, 2026}

\author{
Mustafa Abdool$^{1,*}$ \and
Soumyadip Banerjee$^{1}$ \and
Moutupsi Paul$^{1}$ \and
Do-kyum Kim$^{1}$ \and
Xiaowei Liu$^{1}$ \and
Bin Xu$^{1}$ \and
Tracy Yu$^{1}$ \and
Hui Gao$^{1}$ \and
Karen Ouyang$^{1}$ \and
Huiji Gao$^{1}$ \and
Liwei He$^{1}$ \and
Stephanie Moyerman$^{1}$ \and
Sanjeev Katariya$^{1}$
\\[1ex]
$^{1}$Airbnb, Inc., USA \\
}

\date{January 10, 2026}

\maketitle

\begin{abstract}
The goal of Airbnb search is to match guests with the ideal accommodation that fits their travel needs. This is a challenging problem, as popular search locations can have around a hundred thousand available homes, and guests themselves have a wide variety of preferences. Furthermore, the launch of new product features, such as \textit{flexible date search,} significantly increased the number of eligible homes per search query. As such, there is a need for a sophisticated retrieval system which can provide high-quality candidates with low latency in a way that integrates with the overall ranking stack.

This paper details our journey to build an efficient and high-quality retrieval system for Airbnb search. We describe the key unique challenges we encountered when implementing an Embedding-Based Retrieval (EBR) system for a two sided marketplace like Airbnb---such as the dynamic nature of the inventory, a lengthy user funnel with multiple stages, and a variety of product surfaces. We cover unique insights when modeling the retrieval problem, how to build robust evaluation systems, and design choices for online serving. The EBR system was launched to production and powers several use-cases such as regular search, flexible date and promotional emails for marketing campaigns. The system demonstrated statistically-significant improvements in key metrics, such as booking conversion, via A/B testing.

\end{abstract}

\def\thefootnote{*}\footnotetext{Work done while employed at Airbnb, correspondance to moose878@gmail.com}\def\thefootnote{\arabic{footnote}}

\section{Introduction}
Airbnb search is the main entry point for almost all guests that book an accommodation on the platform. In particular, the goal of the search system is to display the most relevant homes from a potentially large set of eligible homes while taking into account query context and user preferences. Given the high volume of daily searches, the search system must be scalable (scoring millions of homes per second!) while also surfacing relevant results to users with low latency.

Previously, most of the innovations developed for Airbnb search were focused purely on the design of a heavyweight machine learning model (known as the \textit{first-stage-model}) to predict the relevance of each home (henceforth referred as a \textit{listing}) \cite{haldar2020improvingdeeplearningairbnb} \cite{haldar_2019_applying}. This was a reasonable area to focus on when the majority of search queries had a relatively small number of available listings, as each one could be scored individually with acceptable overall latency. 

However, in recent years, new product features which relax constraints on key query parameters such as  \textbf{Flexible Date Search} \cite{airbnb_fds_press_release} and \textbf{Split Stays} \cite{airbnb_split_stays_press_release} led to a significant increase in the number of eligible listings for a typical search query. In addition, with millions of homes available even searches for high demand destination (such as \textit{Paris} or \textit{London}) yielded a significant number of results which put strain on the existing infrastructure.

In light of these changing requirements, we turned our attention towards developing an efficient candidate retrieval system. This system would effectively filter down the large set of eligible homes to a tractable number that could be processed by the more compute-heavy portions of the search ranking stack. The natural choice for such a system was one that leveraged \textit{Embedding-Based Retrieval} (EBR) to represent both listings and search queries as numerical vectors in the same space.

\subsection{Challenges}

While EBR systems have been leveraged in previous industry applications, we encountered several unique challenges when building a retrieval system for a \textit{two-sided e-commerce marketplace} like Airbnb search. 

The first challenge is that Airbnb users generally undergo a \textbf{multi-stage search journey}. The reason is that booking an accommodation on Airbnb is similar to other big-ticket purchases, as the typical user only goes on a few trips a year. As such, users can spend up to several weeks conducting many searches and performing various actions on the platform (such as clicking into listing details, reading reviews, or adding prospective listings to a \textit{wishlist}) before making a final booking. The early stages of a user's search journey tend to be \textit{exploratory} and consist of broad searches with many eligible listings. Hence, from a modeling point of view, ideally the training data for the EBR model should reflect this diverse multi-stage journey and take into account the variety of actions a user can perform.

Once we have a hypothesis on how to model the user's multi-stage journey appropriately, the next challenge lies in evaluation. A two-sided marketplace like Airbnb imposes key constraints on our \textbf{offline evaluation system} - which is an important step to filter ideas before using valuable A/B testing bandwidth. In particular, due to the dynamic nature of Airbnb inventory, it is not computationally feasible to re-construct offline the \textit{entire set} of eligible listings at a specific point in time as hosts are frequently updating pricing and availability information. In addition, it is also not practical from a scalability perspective to log the entire full eligible set of listings during the online request path either. This presents a challenge with offline evaluation using historical logs as they generally only contain the results actually \textit{shown} to a user. As such, metrics obtained by processing such logs (such as \textit{recall@K}) tend not to generalize well to a true online setting.

The final challenge is scaling our retrieval system to meet the demands of real production traffic. A key complication here for platforms like Airbnb is the \textbf{scale of real-time updates}. Unlike traditional Information Retrieval scenarios \cite{mitra2018introduction}, we cannot assume a static corpus of documents over time. In fact, Airbnb listings are constantly being updated by hosts to reflect new information---mostly related to \textit{availability} and \textit{pricing}. These updates are around the scale of \textbf{10,000 per second} and are very sensitive to indexing delays. Furthermore, almost all searches on Airbnb contain some type of filtering criteria (most commonly a geographic region filter if a user moves the map area). Hence, when leveraging indices optimized for Approximate Nearest Neighbor (ANN) computations, we must consider various trade-offs in order to balance latency and performance especially when using such indices in conjunction with real-time updates and traditional search filters.

\subsection{Main Contributions}

The goal of this paper is to describe our journey---in particular, the challenges and design decisions made in  developing the first ever Embedding-Based Retrieval solution for Airbnb search. We present practical learnings that should be applicable to any similar two-sided e-commerce marketplace with regards to these key areas:

\begin{enumerate}
    \item \textbf{Modeling for Multi-Stage Journey:} We provide insight into ways to construct training data for retrieval models given the multi-stage nature of a user's search journey. In addition, we focus on the usage of various actions on the platform to construct \textit{hard negatives} for contrastive learning along with practical trade-offs for model architectures and loss functions.
    \item \textbf{Offline Evaluation:} We describe a novel solution to evaluate models offline by replaying historical traffic to an offline cluster and collecting useful retrieval metrics. One such metric is the recall@100 when our heavyweight first-stage model is treated as the ground truth
    \item \textbf{Online Serving:} We describe the trade-offs made when adopting specialized ANN solutions such as Hierarchical Navigable Small Worlds (HNSW) \cite{malkov2018efficientrobustapproximatenearest} and Inverted File Index (IVF) in the context of the previous scaling challenges (real-time updates and integration with regular search filters). We detail the design of our overall retrieval system and how it was generalized to other product use-cases such as \textit{Flexible Date Search} and \textit{Promotional Emails}.
\end{enumerate}

\section{Search Architecture}
Airbnb search uses a \textit{sharded scatter-gather architecture} to scale horizontally. Each incoming search request first goes to the Root Service, which fans out to make parallel calls to each Leaf Service. Each Leaf Service node performs retrieval and ranking and returns the top results to the Root Service where they are aggregated and further processed.

As with many modern recommender systems \cite{li2021embeddingbasedproductretrievaltaobao}, search ranking at Airbnb does not consist of a single model. At a high-level, there are three main models used throughout the process:

\begin{enumerate}
    \item \textbf{Retrieval/Candidate Selection Model (Leaf Service):} A low-latency model which can evaluate and filter tens of thousands of homes. The top $K$ results are sent to the next stage.
    \item \textbf{First-Stage Model (Leaf Service):} A heavyweight DNN model \cite{haldar2020improvingdeeplearningairbnb} which produces a single score per listing. The top $N$ are sent to the next stage.
    \item \textbf{Setwise Re-Ranker Model (Root Service):} A setwise ranking model to optimize the top $T$ results jointly \cite{abdool2020managingdiversityairbnbsearch}
\end{enumerate}

\begin{figure}[htbp]
  \centering
  \includegraphics[width=\linewidth]{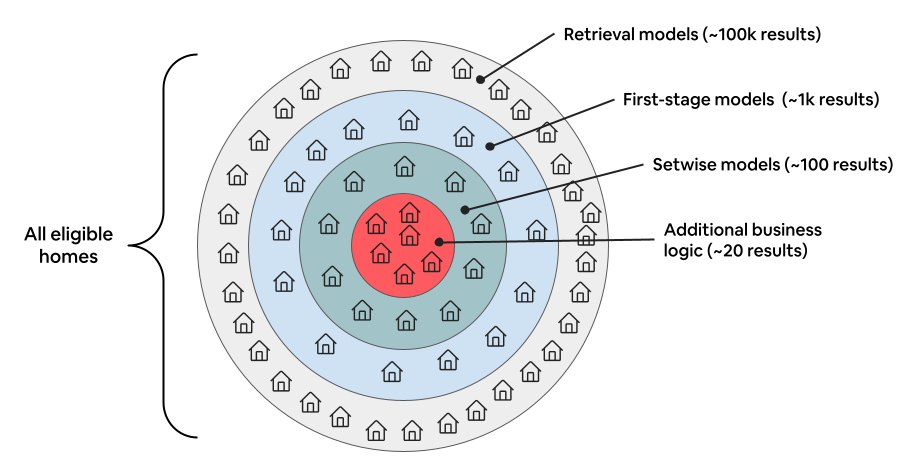}
  \caption{Core ranking models in the Airbnb search ranking stack, each phase scoring an order of magnitude less results than the previous one}
  \label{fig:ranking_stack}
\end{figure}

Note that we have $K > N > T$ where $K$, $N$, $T$ are the maximum numbers of listings processed by the retrieval system, the first stage model, and the setwise model, respectively. Each stage is essentially responsible for scoring an order of magnitude of homes less than the preceding one, resulting in various design trade-offs. This is illustrated in Figure \ref{fig:ranking_stack}. In previous work, we have focused mainly the latter two stages \cite{Tan_2023} as they were most directly amenable to modeling improvements and measurement via A/B testing. However, in recent years, more effort has been spent on optimizing the retrieval system due to increasing scale of search results.

\section{Related Work}

The idea that one can learn dense representations of entities has been one of the major breakthroughs in deep learning over the past decade \cite{bengio2014representationlearningreviewnew}. One of the first popular demonstrations was in \textit{Word2Vec} \cite{mikolov2013efficientestimationwordrepresentations} where it was shown that a learned embedding can transform words into a vector space where conventional distance metrics have semantic meaning. The field has progressively rapidly since then, especially in the domain of Information Retrieval (IR), where embeddings of textual information were used to supplement (or even outperform) traditional methods for the query-document search \cite{mitra2018introduction}.

In the past few years, Embedding-Based Retrieval techniques have also found themselves making an impact in industry. They have been applied to major products such as Facebook Search \cite{Huang_2020} and Taobao Search \cite{li2021embeddingbasedproductretrievaltaobao}. Such papers detail not only the generation of embeddings via techniques such as contrastive learning \cite{chopra_contrastive} but also practical advice on how to tune Approximate Nearest Neighbor (ANN) systems at serving time to achieve practical latency goals. On the modeling side, more recent papers have investigated how to fine-tune and fuse embeddings for pre-trained models such as BERT \cite{lu2020twinbertdistillingknowledgetwinstructured} and end-to-end optimization of the ranking stack \cite{Agarwal_2024}.

\section{Modeling}

\subsection{Training Data Construction}

At the heart of any machine learning model is the training data. We explored two main approaches to create the training data for our retrieval models. The first, \textit{search-based sampling}, was our classic approach for constructing training data and had been used in previous applications \cite{Tan_2023} \cite{haldar2020improvingdeeplearningairbnb}. However, we discovered key limitations to this approach when applying it to retrieval models and we were able to improve upon it with a \textit{trip-based sampling} solution.

\subsection{Search-Based Sampling}

\begin{figure}[htbp]
  \centering
  \includegraphics[width=\linewidth]{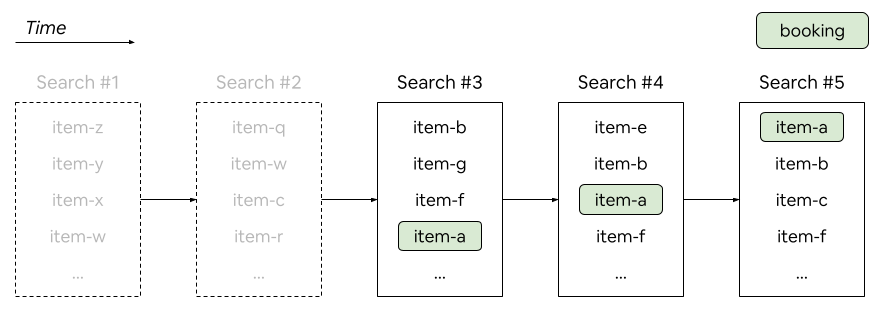}
  \caption{Example of search-based sampling to construct training data. The user makes a booking after doing five searches but their early searches without a booked listing (i.e. Search \#1 and Search \#2) are discarded}
  \label{fig:basic_attribution}
\end{figure}

Our baseline training data construction method involved finding the de-identified booked listing in a user's historical searches (using a fixed look-back window) and assigning it a \textit{positive} label while treating all other impressed listings as \textit{negatives}. Critically, this means that \textit{searches where the booked listing did not appear are discarded}. An example of this scheme is shown in Figure \ref{fig:basic_attribution}.

\subsection{Trip-Based Sampling}

The trip-based sampling approach had two key differences from the search-based one. The first was that we instead grouped together all of a user's searches that had the same important query parameters---such as location, number of guests, and bucketized length-of-stay. We still attributed a positive label to the booked listing, but there was now a much more diverse set of negative listings included overall. This was because we could now include listings from \textit{earlier} searches even when the booked listing was not directly shown in the results.

\begin{figure}[h]
  \centering
  \includegraphics[width=0.7\linewidth]{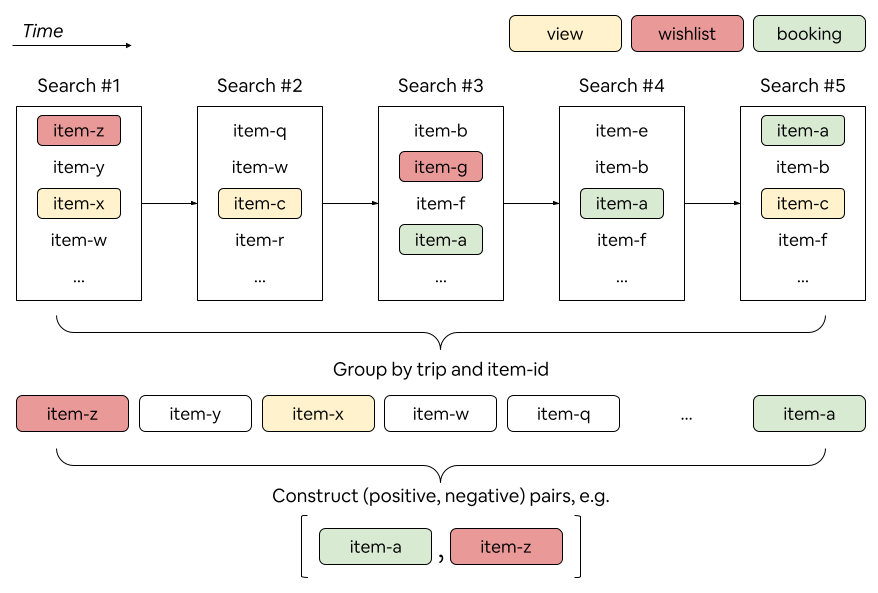}
  \caption{Sampling when done at trip level as used in retrieval training data. Note that early searches in a user's journey are now retained and used to construct contrastive pairs}
  \label{fig:search_journey_tripwise}
\end{figure}

The second key difference was the process of selecting negative examples---in the search-based approach, all listings that were not booked were given equal weight as negatives. This was suboptimal as, when applying contrastive learning, it is well-known that the selection of negative examples is an important factor in robustness and generalization \cite{magnani2024semanticretrievalwalmart}. As such, to add more diversity to our negative sampling population, we assigned \textit{auxiliary labels} based on a variety of \textit{known intentful actions} on the Airbnb platform such as \textit{clicks} and \textit{wishlists} (saves). Then, to sample negative listings for our trip-based method,  we first randomly selected a \textit{category} of actions (impressed, viewed, wishlisted) and then a listing from that category. An example of this trip-based sampling is shown in Figure \ref{fig:search_journey_tripwise}.

\subsection{Comparison of Training Data Sampling}

Overall, we found that trip-based sampling outperformed search-based sampling even when controlling for the overall training data size. In fact, several of our initial experiments for retrieval models leveraged the search-based sampling method (as it was easy to re-use from our first-stage model training data pipeline) but failed to yield any promising results.

By analyzing the data, we realized that the search-based sampling scheme was clearly biased towards the end of a user's search journey, as it requires the booked listing to actually be \textit{shown} in the results. From analyzing the data, we found that if we temporally ordered all searches by booked users, this scheme resulted in the majority of listings being sampled from the \textbf{last 30\%} of the journey. As such, by performing trip-based sampling, we were able to mitigate this temporal bias by including a more diverse (and harder) set of listings that appeared early on. This enabled the model trained on trip-based sampling data to generalize better across \textit{all} stages of a user's search journey.

\begin{figure}[h]
  \centering
  \includegraphics[width=0.6\linewidth]{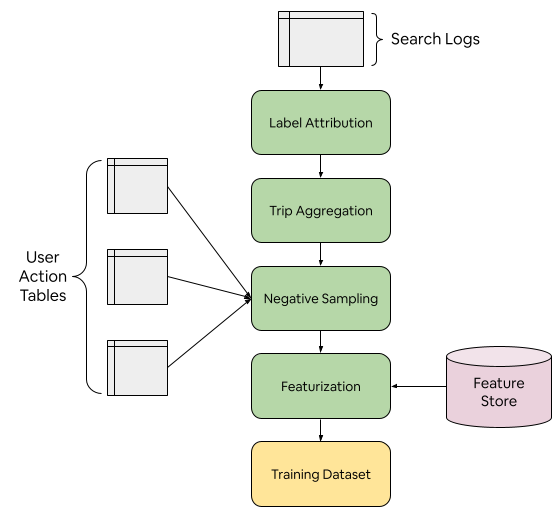}
  \caption{Training data pipeline for the trip-based sampling with harder negatives as used in retrieval model training }
  \label{fig:trip_based_attribution_pipeline}
\end{figure}

Another key factor was the selection of negative listings. In search-based sampling, all impressed listings had an equal chance of being a negative, even though users generally only engage with a few listings per page. To rectify this, we introduced the auxiliary labels strategy for sampling negatives described in previous section - which generated much \textit{harder negatives} as it up-sampled listings for which a user took more intentful actions. An overview of the final data pipeline for retrieval modeling is shown in Figure \ref{fig:trip_based_attribution_pipeline}.

\subsection{Model Architecture}

\begin{figure}[htbp]
  \centering
  \includegraphics[width=\linewidth]{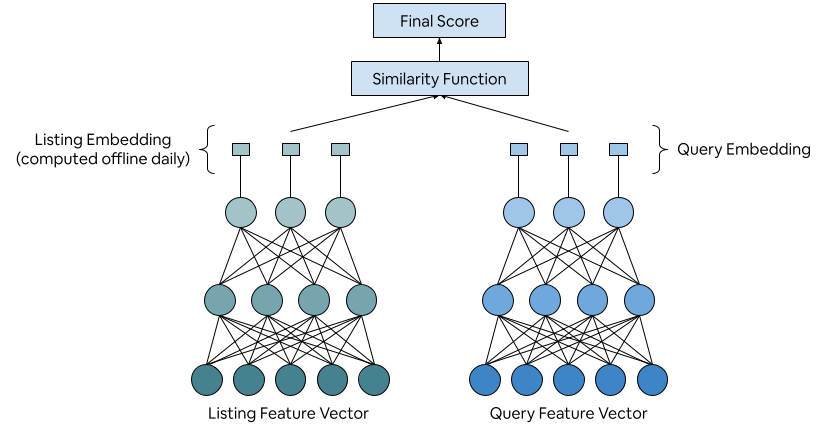}
  \caption{Two-tower architecture as used in EBR model. Note that the listing tower is computed offline daily}
  \label{fig:simplehead_arch}
\end{figure}

In terms of model architecture, we used two-tower model commonly seen in industry applications and that we had leveraged in the past \cite{haldar2020improvingdeeplearningairbnb}. The final model output ($F_{\theta}$) is computed as:

\begin{equation}
  F_{\theta}(l,q) = \text{sim}(Q_{\theta}(q), L_{\theta}(l))  
\end{equation}

where $l$ and $q$ are feature representations of a listing and query respectively. The listing tower ($L_{\theta}$) and the query tower ($Q_{\theta}$) are deep neural networks which learn embeddings for the query and listing respectively which are then combined. The embeddings are combined via a similarity function ($sim$), such as \textit{Euclidean distance} or even a \textit{dot product}. The overall architecture is shown in Figure \ref{fig:simplehead_arch}.

A key technical decision is that we designed the features used in the listing tower such that it could be computed entirely in an \textit{daily offline batch job}. In this way, we only need to compute the query tower at inference time, which greatly reduced the compute requirements as it \textit{no longer scales with the number of eligible listings}. However, this did come with the trade-off of being unable to use real-time listing features initially. For example, the previous day's price and availability features are used in the initial retrieval phase and then the up-to-date price and availability information are refreshed and used later in the search process.

In terms of listing features used, we perform aggregations based on historical engagement such as past views, wishlists, and reviews. We further include a large amount of features that are engagement-independent such as amenities, location (\textit{S2Cell} IDs of various resolutions), and guest capacity to mitigate the cold start problem and generalize across the diverse population of listings.

With regards to query features, we try to keep the set small in order to reduce the online inference cost of the query tower. Key information such as the location (encoded via a canonical \textit{placeId} and represented by a learned dense embedding), number of guests, and geographic size of the search location is used.

\section{Offline Evaluation}

For many two-sided marketplaces, online A/B testing traffic is considered very valuable which means that having a trustworthy offline evaluation pipeline is essential to iterating quickly on ideas. As mentioned previously, evaluating a retrieval model offline can be quite challenging as they are most impactful on search queries with a large number of listings (ie. tens of thousands) but yet we cannot fully log all results for every search due to system constraints.

\subsection{Logged Data with Random Negative Listing Samples}

While offline evaluation via historical logs is not perfect, it can still be a useful metric to guide general model development. We evaluate retrieval models on a held-out set of search data that occurs \textit{after} the training period. We also made a modification to our online logging system to sample a small random negative samples of available but \textit{unshown} listings for each search query. The intuition here is that such negative  approximately simulates the retrieval case where a large number of diverse listings need to be scored. We compute the \textbf{recall} (as opposed to NDCG which is better suited to evaluate the first-stage model) of the booked listing at various thresholds.

\subsection{Traffic Replay Framework}
As mentioned as one of the initial challenges, a key limitation of re-scoring the model on offline logs is that the number of results is \textit{orders of magnitude less} than the type of searches a retrieval model would actually be used for online. In order to have our offline evaluation match the scale of results retrieved in production, we used a \textit{ traffic replay framework} to fork a small percentage of production search traffic to an offline cluster. As this is an offline cluster, we do not need to be concerned with latency, so we can modify the request to score \textit{all eligible listings} using \textit{both} our heavyweight first-stage model and a candidate retrieval model.

\begin{figure}[htbp]
  \centering
  \includegraphics[width=\linewidth]{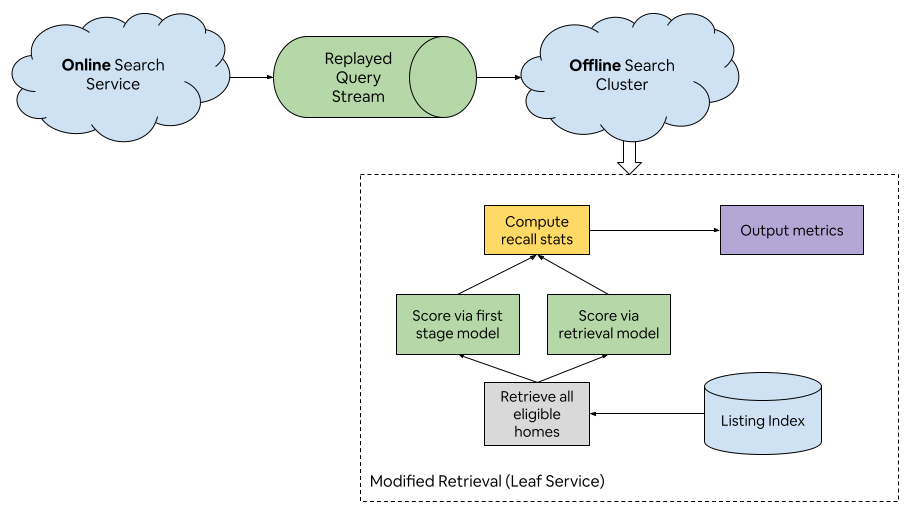}
  \caption{Framework to replay real search queries and compute recall metrics for a retrieval model using the first-stage model as the ground truth (runs in offline cluster)}
  \label{fig:offline_evaluation_framework}
\end{figure}

We then computed the recall of our retrieval model at various thresholds, by treating the top $K$ results ordered by our first-stage model score as the ``ground truth'' positives (where $K$ is much smaller than the total number of listings available). The reasoning here is that we generally want our retrieval model to surface all listings that would have had a high first-stage model score---so this becomes a useful proxy measure for how well our retrieval model is performing. Empirically, we found this metric was predictive of the final A/B test performance for all candidate retrieval models ($\approx 10)$ that we tested during this work. A diagram of this system is shown in Figure \ref{fig:offline_evaluation_framework}.

\subsection{ANN System Design Considerations}

Our two main candidates for online ANN system were Inverted File Index and Hierarchical Navigable Small Worlds (HNSW) \cite{malkov2018efficientrobustapproximatenearest}---both popular solutions in industry. We benchmarked several approaches using the FAISS library \cite{douze2024faiss} and found that, overall, IVF had the best trade-off between speed and performance for our use case.

As mentioned previously, unique aspects of our case are that nearly all Airbnb searches are issued with some type of filter (especially a geographic region filter if the user interacts with the map) and a high-volume of realtime updates for listing data. We observed HNSW suffered from poor performance when doing filtering in conjunction with the high number of realtime updates per listing: The index size and memory footprint often grew too large. Updating the HNSW graph at a high QPS was problematic in the Airbnb implementation of HNSW as well which meant that discriminative filters were difficult to apply in parallel with retrieval without major system changes.

In contrast, IVF only requires cluster IDs and centroids to be stored in the index at runtime, allowing much lower memory usage and the ability to treat them as a normal filter in Apache Lucene \cite{ApacheLucene}. The trade-off, however, was that IVF yielded poorer recall overall than HNSW on average. Ultimately, we decided to use IVF in order to maintain system stability and integrate easily with our current search indexing pipeline.

\subsection{IVF System Details}

\begin{figure}[htbp]
  \centering
  \includegraphics[width=0.7\linewidth]{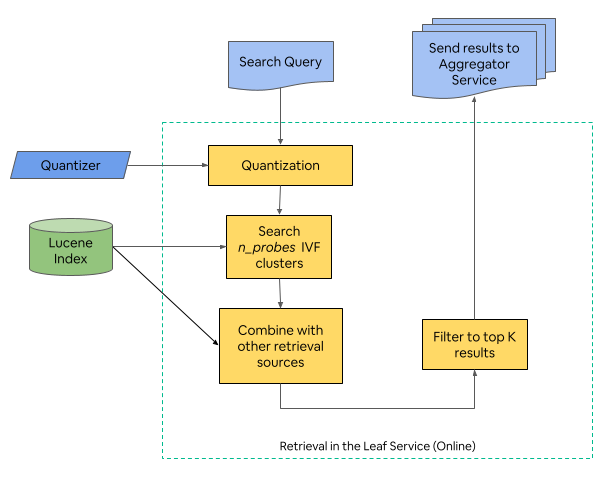}
  \caption{Full serving stack for the EBR model and IVF. Listings are clustered daily and cluster ID tags are built into the index itself to be used at serving time}
  \label{fig:full_serving_stack}
\end{figure}

Given our choice to use IVF, we introduced an offline \textit{k-means clustering} stage in our offline data pipeline that ran daily after the listing embeddings are re-computed; the cluster IDs for each listing were then stored as part of the listing index. A diagram for this full serving solution is shown in Figure \ref{fig:full_serving_stack}. One challenge was how to determine the value of an important parameter known as \textit{nprobes} which defines the number of closest clusters (to the \textit{query embedding}) to retrieve listings from. This \textit{nprobes} parameter is critical to control performance versus latency trade-offs, as a higher value will result in better recall but at the cost of higher compute and latency. 

\begin{figure}[htbp]
  \centering
  \includegraphics[width=0.7\linewidth]{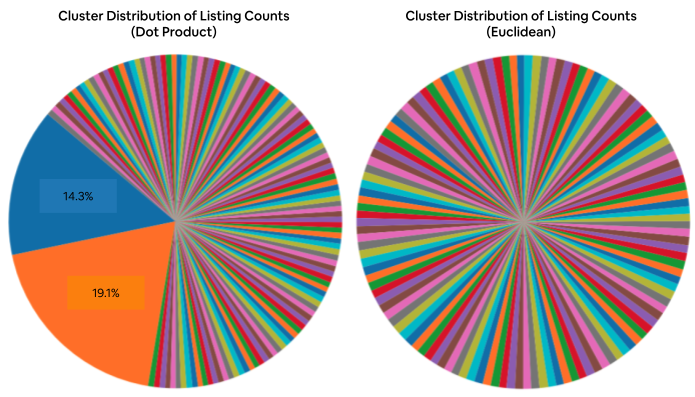}
  \caption{Distribution of cluster sizes (128 total) when using dot product vs. Euclidean distance. Dot product frequently resulted in 1\textendash2 outlier cluster sizes that contained $\sim$20\% of the data}
  \label{fig:cluster_distributions}
\end{figure}

To determine the optimal value of \textit{nprobes} we wrote an offline simulation framework to compare the recall of true KNN retrieval vs. IVF for various values of \textit{nprobes} and selected a value to balance recall and compute costs. This led us to an interesting finding which was that certain similarity functions resulted in much worse clustering than others! Specifically, even though \textit{dot product} and \textit{Euclidean distance} performed about the same in offline evaluation, we found that \textit{dot product} similarity produced clusters with very imbalanced sizes. This was even after a normalizing transformation from maximizing the dot product into the Euclidean nearest neighbor problem \cite{msftInnerProductPaper}---something others had reported as well \cite{magnani2024semanticretrievalwalmart}. However, when we switched to using a \textit{Euclidean} distance, the clusters were much more balanced and a much lower value of \textit{nprobes} was required. We suspect this occurs because dot product similarity inherently does not take the magnitude of the embedding values themselves (only the direction) but many of the underlying features are derived from numerical counts (such as engagement statistics). See Figure \ref{fig:cluster_distributions} for a visualization of the cluster size distributions.

\subsection{Additional Product Use Cases}

\subsubsection{Flexible Date Search}

A further challenge for our EBR system was scaling to support use cases such as \textit{Flexible Date Search} \cite{airbnb_fds_press_release} which greatly expanded the number of eligible results. For example, if a guest is flexible over +/- 3 days, the checkin/checkout dates can have 49 unique combinations---a naive solution would involve issuing 49 separate queries to the EBR system! For this case, the overall retrieval system was bottle-necked by the expensive computation to determine if a listing was available (and at what price) for all subsets of a date range.

To handle this, we created a compact \textit{data structure} of pre-computed availability and pricing \cite{fds_patent} for all listings for a predetermined number of lookahead days in the future. This data structure was also designed with the ability to support near-realtime updates when a listing's data changes and is used during the retrieval phase in conjunction with the EBR model. We have a final post-filtering step, where the accurate pricing and availability checks are refreshed and used before the results are returned from the leaf service.

\subsubsection{Promotional Email Campaigns}

We also were able to generalize our EBR system to further support use cases across the company such as from the \textit{Marketing Technology} division. For example, generating listings for \textit{promotional email campaigns} involves issuing searches with specific criteria (ie. a similar price range to a listing a user had considered booking, but which is now unavailable) but potentially over a broad range of dates and geographic area. As such, we developed a generalized platform for internal services to issue search requests that leveraged both EBR and traditional search filters (ie. find listings similar to a candidate listings and within 10 miles). The batch system which generated promotional emails was then able to leverage this platform to greatly improve the relevance of our email recommendations.

\section{Results}

\subsection{Experiments}

We tested three main variants of our retrieval system:

\begin{enumerate}
    \item \textbf{Baseline:} Our legacy system which uses a simple linear model to generate a \textbf{query-independent score} for all listings daily.
    \item \textbf{EBR V1:} Our first system which uses the two-tower EBR model described previously with a pointwise loss function and inner product similarity. We do not use any ANN solutions here and instead take a union of the top results from the baseline model and the new EBR model scores.
    \item \textbf{EBR V2:} Same strategy as the V1 system but the EBR model now uses an IVF solution to reduce runtime and compute costs.
    \item \textbf{EBR V3:} Uses the same IVF strategy as in the V2 system but with an updated EBR model which now includes more features, larger network size (20\% more parameters) and a pairwise loss function with Euclidean distance.
\end{enumerate}

\subsection{Offline Evaluation}

Table \ref{tab:offline_experiment_results} describes the model performance in terms of recall for the booked listing on a common test set of search data. In addition, Table \ref{tab:offline_experiment_results} shows the recall results from the evaluation via traffic replay discussed in the previous sections also on a common held-out testing date range. We use a higher threshold for this metric as online searches generally retrieve an orders of magnitude more results than those just seen by the user. In both cases, we see that there is a sizable increase in going from the baseline system to incorporating an EBR model and when enhancing the model further.

\begin{table}[ht]
\centering
\caption{Recall for the models of the various retrieval systems (without any ANN techniques used) as measured in offline logs (recall @ 10) and replayed searches (recall @ 100)}
\label{tab:offline_experiment_results}
\begin{tabular}{lcc} % l = left, c = center (column alignment)
\toprule
Experiment & Recall (Offline Logs) & Recall (Replayed) \\
\midrule
Baseline & 53.3\% & 40.5\% \\
EBR Model (V1 and V2) & 60.2\% & 74.6\% \\
\textbf{EBR Model (V3)} & \textbf{93.4\%} & \textbf{87.0\%} \\
\bottomrule
\end{tabular}
\end{table}

\subsection{Online A/B Testing Results}

We conducted online A/B tests for each of the three major versions mentioned above. In total, the successive launches of these three retrieval systems yielded a combined relative \textbf{0.31\%} increase in conversion gains and the V3 system is currently in production. This conversion gain is a large portion of the team's annual goal and is on par with recent major innovations (\cite{Tan_2023}
\cite{abdool2020managingdiversityairbnbsearch} \cite{haldar2023learningrankdiverselyairbnb}). In addition, when adopting IVF as an ANN solution in the V2 system, we were able to yield a reduction in compute resources of around \textbf{16\%} relative to the baseline system.

As an added benefit, we saw a statistically-significant increase in bookings of \textit{new listings} on the platform along with an increase in bookings from \textit{wishlisted} listings. These new listing booking gain highlights the fact that the EBR model was able to generalize much better than the baseline - likely due to having access to features which repersent the query context rather than having to rely on historical engagement data. The increase in wishlisted bookings provides evidence to the theory that our EBR system was surfacing more useful listings early in the user's journey (due to the trip-based training data) as it is known that users tend to wishlist potential listings early on.

Lastly, for \textit{email promotional campaigns} specifically, our A/B test yielded +2.3\% bookings from email promotions, indicating the proposed approach is able to retrieve more relevant listings that users prefers to booking when they receive promotional emails.

\section{Key Lessons and Takeaways}

\begin{enumerate}
    \item \textbf{Importance of Negative Candidate Selection:} The core idea of the \textit{trip-based} aggregation stemmed from including a more diverse set of negative listings during model training to better represent the user search journey. This was a key insight to reduce bias and allow the retrieval model to generalize to a more diverse set of listings and search queries.
    \item \textbf{Accurate Offline Evaluation Systems:} In practice, bandwidth for A/B testing can be quite limited due to the company-wide queue of potential changes. As such, it is key to build robust offline evaluation systems beyond just relying on offline logging, which usually does not fully represent the true distribution of eligible listings. We were able to solve this problem by using a novel traffic replay framework to spend more compute in order to generate retrieval metrics that were more predictive of online experiments
    \item \textbf{Choosing an ANN System:} While we were initially excited about HNSW, we ultimately found it did not work well for our traffic patterns which involved many real-time updates and a high proportion of searches with discriminative filters. As such, the takeaway here is to consider the query distribution and existing system architecture when choosing an ANN system. Even once we decided to use IVF, it was important to add monitoring to debug situations such as when Euclidean distance produced imbalanced clusters and identify performance gaps
\end{enumerate}

\section{Extensions and Future Work}

While we have made great strides into bringing EBR to Airbnb search, there is still much future work to explore. In the near future, we plan to explore a variety of new retrieval models beyond the two mentioned in this paper (the baseline query-independent model and the query-dependent EBR model), such as:

\begin{enumerate}
    \item \textbf{Retrieval Model with Past Context:} For listings which the user has appeared to interact with in the past (such as through \textit
    {clicks} or adding to various \textit{wishlists}).
    \item \textbf{Geo-Intent Retrieval Model:} For broad area searches that encompass multiple major cities (such as \textit{California} or \textit{France}), determine which cities would be relevant for that area.
\end{enumerate}

\begin{figure}[htbp]
  \centering
  \includegraphics[width=\linewidth]{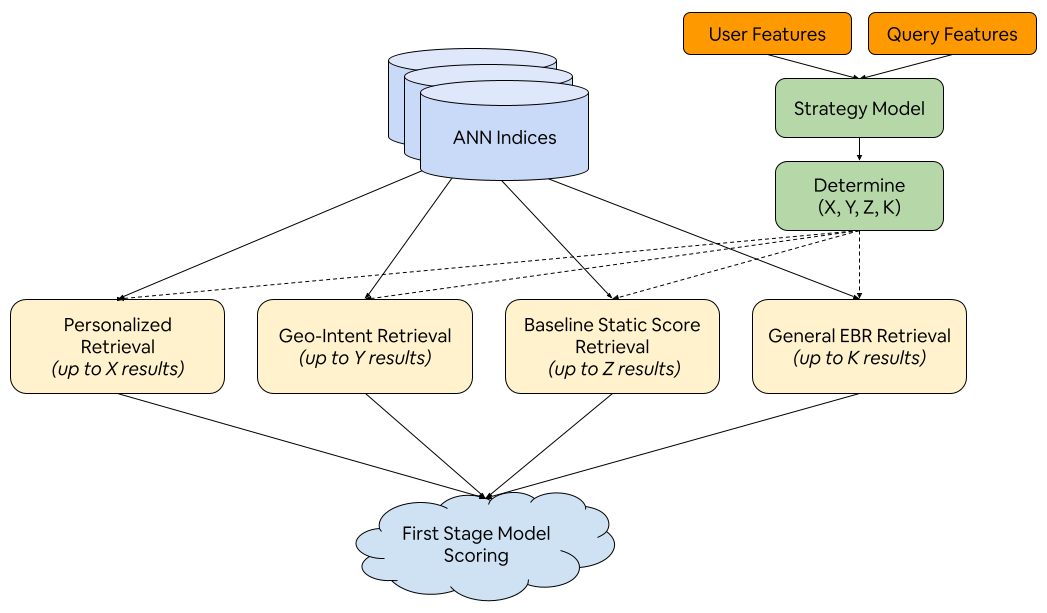}
  \caption{Architecture for a generalized retrieval system that can support multiple sources and a strategy model which delegates the relative importance (number of listings) to retrieve from each source}
  \label{fig:future_retrieval_system}
\end{figure}

With various types of retrieval models, we can then build a \textit{strategy layer} to dynamically allocate retrieval limits for each one. A diagram of this overall system is shown in Figure \ref{fig:future_retrieval_system} and a new strategy layer model is currently under experimentation with promising initial results.

\section{Conclusion}

Overall, the journey to apply Embedding-Based Retrieval to Airbnb search has been a fruitful one but certainly not without its own challenges. Foremost among them include deeply understanding the retrieval model problem given a multi-stage user search journey, designing an offline evaluation system that enables fast iteration cycles, and building a serving system to meet the requirements of an online platform with millions of daily active users. 

Most importantly, we are grateful for the support of our friends, colleagues, and the broader machine learning community---and look forward to tackling new challenges on the horizon together.

% \begin{acks}
% We would like to specifically thank Yangbo Zhu and the entire Search Infrastructure team at Airbnb. Without their great collaboration this paper would not be possible!
% \end{acks}

% to move onto the next page
\clearpage

\appendix

\section{Appendix: IVF with different similarity measures}

Table \ref{tab:sim_comparision} shows the recall of IVF vs. true KNN as a function of $nprobes$. Here we see the importance of the Euclidean distance in producing balanced clusters as it greatly reduces the $nprobes$ needed to achieve high recall.

\begin{table}[ht]
\centering
\caption{Recall with IVF vs. True KNN for different similarity measures (128 clusters total) when using the model from EBR V3 on a sample of query embeddings. Euclidean distance greatly outperforms dot product due its ability to produce more balanced clusters}
\label{tab:sim_comparision}
\begin{tabular}{lcc} % l = left, c = center (column alignment)
\toprule
Similarity Measure & n probes & recall @ 100 vs. true KNN \\
\midrule
Dot Product & 2 & 43\% \\
Dot Product & 4 & 47\% \\
Dot Product & 16 & 66\% \\
Dot Product & 32 & 84\% \\
Euclidean Distance & 2 & 71\% \\
Euclidean Distance & 4 & 75\% \\
\textbf{Euclidean Distance} & \textbf{16} & \textbf{98\%} \\
Euclidean Distance & 32 & 100\% \\
\bottomrule
\end{tabular}
\end{table}

\section{GenAI Usage Disclosure}

No GenAI tools were used in the process of writing this paper.

\bibliographystyle{plain}
\bibliography{sample-base}

\end{document}